\documentclass[11pt]{article}
\usepackage{moriond}
\usepackage{epsfig}
\usepackage{graphicx}
\usepackage{color}

\def\Journal#1#2#3#4{{#1} {\bf #2}, #3 (#4)}

\def\be{\begin{equation}}
\def\ee{\end{equation}}
\def\bea{\begin{eqnarray}}
\def\eea{\end{eqnarray}}

\newlength {\picwidth}
\newcommand{\cerenkov}  {\v{C}erenkov}   
\newcommand{\DO}        {\mbox{\ensuremath{{\rm D}_2{\rm O}}}} 
\newcommand{\HO}        {\mbox{\ensuremath{{\rm H}_2{\rm O}}}} 
\newcommand{\B}         {\mbox{\ensuremath{^{8}{\rm B}}}}
\newcommand{\iB}        {\mbox{\ensuremath{^{\it 8}\!{\it B}}}}
\newcommand{\N}         {\mbox{\ensuremath{^{16}{\rm N}}}}
\newcommand{\Clf}       {\mbox{\ensuremath{^{35}{\rm Cl}}}}
\newcommand{\Cls}       {\mbox{\ensuremath{^{36}{\rm Cl}}}}
\newcommand{\Natt}      {\mbox{\ensuremath{^{23}{\rm Na}}}} 
\newcommand{\Natf}      {\mbox{\ensuremath{^{24}{\rm Na}}}} 
\newcommand{\iNatf}     {\mbox{\ensuremath{^{\it 24}\!{\it Na}}}} 
\newcommand{\nue}       {\mbox{\ensuremath{\nu_{e}}}}
\newcommand{\numu}      {\mbox{\ensuremath{\nu_{\mu}}}}
\newcommand{\nutau}     {\mbox{\ensuremath{\nu_{\tau}}}}

\newcommand{\cm}        {\mbox{\,cm}} 
\newcommand{\m}         {\mbox{\,m}}
\newcommand{\ns}        {\mbox{\,ns}}
\newcommand{\Hz}        {\mbox{\,Hz}}
\newcommand{\MeV}       {\mbox{\,MeV}} 
\newcommand{\Tl}        {\mbox{\ensuremath{^{208}{\rm Tl}}}}
\newcommand{\Bi}        {\mbox{\ensuremath{^{214}{\rm Bi}}}}

\begin{document}
\vspace*{4cm}  
\title{FIRST RESULTS FROM THE SUDBURY NEUTRINO OBSERVATORY}
\author{G.A.\,McGREGOR\,\footnote{ {\scriptsize
for the SNO Collaboration: Q.R.~Ahmad, R.C.~Allen, T.C.~Andersen, J.D.~Anglin,
G.~B\"uhler, J.C.~Barton,  E.W.~Beier, M.~Bercovitch, J.~Bigu
% ...
M.~Yeh.}}}
\address{Department of Physics, Denys Wilkinson Building, Keble Road,
Oxford OX1 3RH, U.K.}

\maketitle\abstracts{
The Sudbury Neutrino Observatory (SNO) is a water imaging \cerenkov\ detector.
Utilising a 1 kilotonne ultra-pure \DO\ target, it is the first experiment to
have equal sensitivity to all flavours of active neutrinos. This allows a
solar-model independent test of the neutrino oscillation hypothesis to be made.
Solar neutrinos from the decay of \B\ have been detected at SNO by the
charged-current (CC) interaction on the deuteron and by the elastic scattering
(ES) of electrons. While the CC interaction is sensitive exclusively to \nue,
the ES interaction has a small sensitivity to \numu\ and \nutau. In this paper,
the recent solar neutrino results from the SNO experiment are presented. The
measured ES interaction rate is found to be consistent with the high precision
ES measurement from the Super-Kamiokande experiment. The \nue\ flux deduced from
the CC interaction rate in SNO differs from the Super-Kamiokande ES measurement
by 3.3$\sigma$. This is evidence of an active neutrino component, in addition to
\nue, in the solar neutrino flux. These results also allow the first
experimental determination of the active \B\ neutrino flux from the Sun, and
this is found to be in good agreement with solar model predictions.}

\section{Introduction}

Over the past 30 years, solar neutrino experiments~\cite{cl,sage,gallex,gno,kam,sk} have
measured fewer neutrinos than are predicted by models of the Sun.~\cite{BPB,TC} A comparison
of the predicted and observed solar neutrino fluxes for these experiments are shown in
table~\ref{intro_fluxes}. These observations can be explained if the solar models are
incomplete or neutrinos undergo a flavour changing process while in transit to the Earth, the
most accepted of which is neutrino oscillations. This puzzle is known as the solar neutrino
problem.

\begin{table}[!t]
\caption{Summary of solar neutrino observations at different solar neutrino
detectors.}
\vspace{0.4cm} 
\label{intro_fluxes}
\begin{center}
\begin{tabular}{|l|l|l|}
\hline
Experiment & Measured Flux & SSM Flux~\cite{BPB} \\ \hline
Homestake~\cite{cl} & 2.56$\pm$0.16({\small \it stat.})$\pm$0.16({\small \it sys.})\,SNU & 7.6$^{+1.3}_{-1.1}$\,SNU \\ \hline
SAGE~\cite{sage} & 67.2$^{+7.2}_{-7.0}$({\small \it stat.})$^{+3.5}_{-3.0}$({\small \it sys.})\,SNU & 128$^{+9}_{-7}$\,SNU  \\ \hline
Gallex~\cite{gallex} &  77.5$\pm$6.2({\small \it stat.})$^{+4.3}_{-4.7}$({\small \it sys.})\,SNU & 128$^{+9}_{-7}$\,SNU  \\ 
GNO~\cite{gno} & 65.8$^{+10.2}_{-9.6}$({\small \it stat.})$^{+3.4}_{-3.6}$({\small \it sys.})\,SNU & 128$^{+9}_{-7}$\,SNU  \\ \hline
Kamiokande~\cite{kam} & 2.80$\pm$0.19({\small \it stat.})$\pm$0.33({\small \it sys.})$\times$10$^6$\,cm$^{-2}$\,s$^{-1}$ &
5.05(1$^{+0.20}_{-0.16}$)$\times$10$^6$\,cm$^{-2}$\,s$^{-1}$ \\
Super-Kamiokande~\cite{sk} & 2.32$\pm$0.03({\small \it stat.})$^{+0.08}_{-0.07}$({\small \it sys.})$\times$10$^6$\,cm$^{-2}$\,s$^{-1}$
& 5.05(1$^{+0.20}_{-0.16}$)$\times$10$^6$\,cm$^{-2}$\,s$^{-1}$ \\ \hline
\end{tabular}
\end{center}
\end{table}

\section{The Sudbury Neutrino Observatory}

\subsection{The SNO Detector}

\begin{figure}[!t]
\begin{center}
\includegraphics[width=\picwidth]{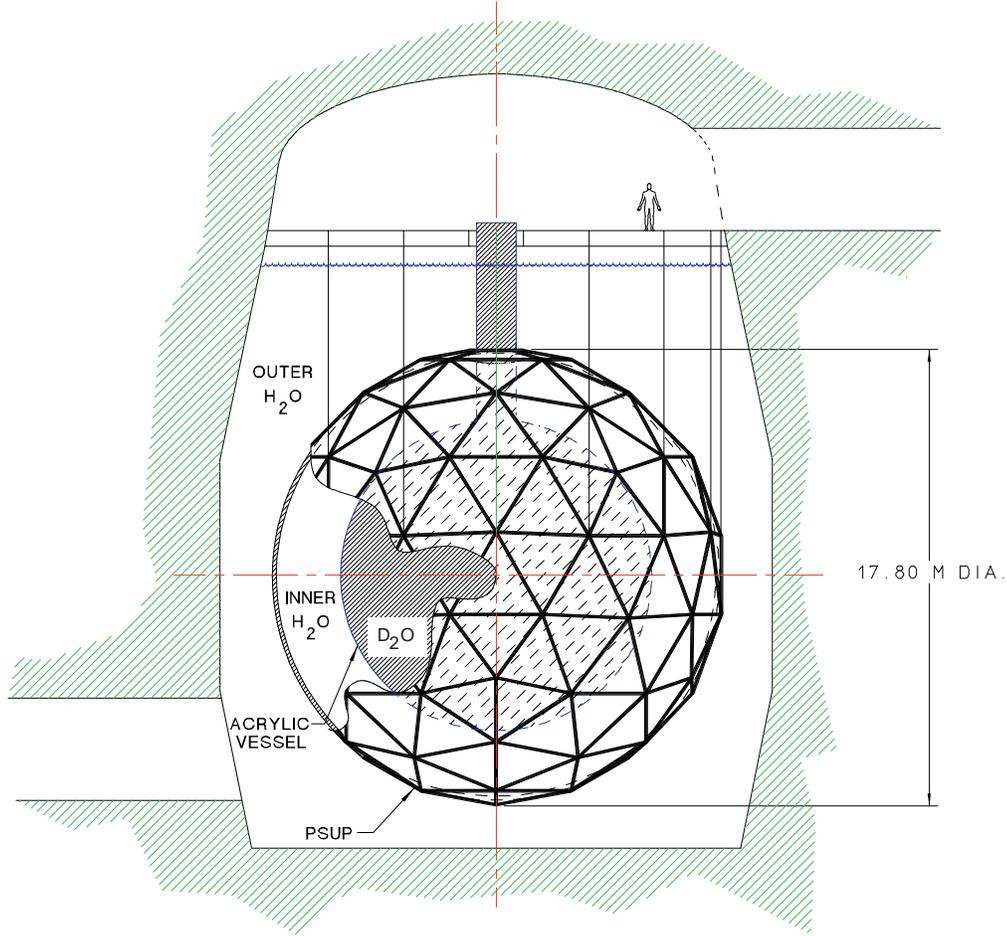}
\caption{A cross-sectional view of the SNO detector.}
\label{sno_detector}
\end{center}
\end{figure}

SNO~\cite{NIM} is an imaging water \cerenkov\ detector located at a depth of
2092\m\ (6010\m\ of water equivalent) in the INCO, Ltd.\ Creighton mine near
Sudbury, Ontario. The detector, shown in figure~\ref{sno_detector}, is situated in
a large barrel shaped cavity 22\m\ in diameter and 34\m\ in height. The 1 kilotonne
ultra-pure \DO\ target is contained within a transparent acrylic vessel (AV) 12\m\
in diameter and 5.5\cm\ thick. A 17.8\m\ diameter geodesic sphere (PSUP -
photomultiplier support structure) surrounds the AV and supports 9456 inward
looking and 91 outward looking 20\cm\ photomultiplier tubes (PMTs). The PSUP is
supported by steel ropes attached to the deck. The remaining volume is filled with
ultra-pure \HO\ which acts as a cosmic ray veto and as a  shield from naturally
occurring radioactivity in both the construction materials and the surrounding
rock. The light water also supports the \DO\ and AV with the remaining weight
supported by 10 Vectran rope loops.

A physics event trigger is generated in the detector when 18 or more PMTs exceed a
threshold of $\sim$0.25 photo-electrons within a coincidence time window of 93\ns.
The trigger reaches 100\% efficiency when the PMT multiplicity is $\ge$23. The
instantaneous trigger rate is about 15-20\Hz, of which 6-8\Hz\ are physics triggers
and the rest are diagnostic triggers.

\subsection{Neutrino Interactions in SNO}

By utilising a \DO\ target, the SNO detector is capable of simultaneously
measuring the flux of electron type neutrinos and the total flux of all active neutrinos
from \B\ decay in the Sun through the following interactions:
\[ \begin{array}{lcll}
\makebox[2cm]{}\nu_e + {\rm d} & \rightarrow & {\rm p} + {\rm p} + e^- & \makebox[2cm]{}(CC)\\
\makebox[2cm]{}\nu_x + {\rm d} & \rightarrow & \nu_x + {\rm p} + {\rm n} & \makebox[2cm]{}(NC)\\
\makebox[2cm]{}\nu_x + e^- & \rightarrow & \nu_x + e^- & \makebox[2cm]{}(ES)
\end{array} \]
The charged-current (CC) interaction on the deuteron is sensitive exclusively to \nue,
and the neutral-current (NC) interaction has equal sensitivity to all active
neutrino flavours ($\nu_x$, {\it x=e,$\mu$,$\tau$}). Elastic scattering (ES) on the
electron is also sensitive to all active flavours, but has enhanced sensitivity to
\nue.

\section{Results from SNO}

The results presented here are the recent results from the SNO
collaboration.~\cite{SNO_PRL1} Full details of the analysis will not be presented
here; readers are encouraged to consult the original paper. The results are from
data recorded between Nov.\ 2, 1999 and Jan.\ 15, 2001, corresponding to 240.95
days of live time. The neutrino fluxes deduced from the CC and ES interactions at SNO
are:
\begin{eqnarray*}
\Phi^{\rm CC}_{\rm SNO} & = &
1.75\pm 0.07({\it stat.})^{+0.12}_{-0.11}({\it sys.})\pm 0.05({\it theor.}) 
\times 10^6\,{\rm cm^{-2}\,s^{-1}} \\
\Phi^{\rm ES}_{\rm SNO} & = &
2.39\pm 0.34({\it stat.})^{+0.16}_{-0.14}({\it sys.})
\times 10^6\,{\rm cm^{-2}\,s^{-1}}
\end{eqnarray*}
where the theoretical uncertainty is the CC cross section
uncertainty.~\cite{xsect} The difference between $\Phi^{\rm CC}_{\rm SNO}$ and
$\Phi^{\rm ES}_{\rm SNO}$ is 0.64$\pm$0.40$\times$10$^6$cm$^{-2}$\,s$^{-1}$, or
1.6$\sigma$. The ratio of $\Phi^{\rm CC}_{\rm SNO}$ to the predicted \B\ solar
neutrino flux given by the BPB01 solar model~\cite{BPB} is 0.347$\pm$0.029 where all
the uncertainties are added in quadrature.

The Super-Kamiokande~\cite{sk} experiment has made a high precision measurement of
the \B\ solar neutrino flux deduced from the ES interaction:
\begin{eqnarray*}
\Phi^{\rm ES}_{\rm SK} & = &
2.32\pm 0.03({\it stat.})^{+0.08}_{-0.07}({\it sys.}) 
\times 10^6\,{\rm cm^{-2}\,s^{-1}} 
\end{eqnarray*}
The measurements $\Phi^{\rm ES}_{\rm SNO}$ and $\Phi^{\rm ES}_{\rm SK}$ are consistent.
Assuming that the systematic errors are normally distributed, the difference between
$\Phi^{\rm CC}_{\rm SNO}$ and $\Phi^{\rm ES}_{\rm SK}$ is
0.57$\pm$0.17$\times$10$^6$cm$^{-2}$\,s$^{-1}$, or 3.3$\sigma$. The probability that
$\Phi^{\rm CC}_{\rm SNO}$ is a $\ge$3.3$\sigma$ downward fluctuation is 0.04\%.

The CC energy spectrum was also extracted from the data and no evidence for spectral
distortions was found.

\subsection{Systematic Uncertainties}
 
\begin{table}[!t]
\caption{Systematic uncertainties on fluxes.}
\vspace{0.4cm} 
\label{sys_errs}
\begin{center}
\begin{tabular}{|l|r|r|}
\hline
Error source  & CC  error  & ES  error\\
                & (percent) & (per cent)\\
\hline
Energy scale            & -5.2, +6.1  & -3.5, +5.4 \\
Energy resolution       & $\pm$0.5       & $\pm$0.3       \\
Energy scale non-linearity           & $\pm$0.5       & $\pm$0.4 \\
Vertex accuracy         & $\pm$3.1       & $\pm$3.3       \\
Vertex resolution       & $\pm$0.7       & $\pm$0.4       \\
Angular resolution      & $\pm$0.5       & $\pm$2.2       \\
High energy $\gamma$'s  & -0.8, +0.0     & -1.9, +0.0  \\
Low energy background   &  -0.2, +0.0     & -0.2, +0.0 \\
Instrumental background &  -0.2, +0.0    & -0.6, +0.0 \\
Trigger efficiency      &  0.0           & 0.0 \\
Live time               & $\pm$0.1       & $\pm$0.1       \\
Cut acceptance          & -0.6, +0.7     & -0.6, +0.7  \\
Earth orbit eccentricity & $\pm$0.1      & $\pm$0.1 \\
$^{17}$O, $^{18}$O      &  0.0           &  0.0 \\
\hline
Experimental uncertainty  & -6.2, +7.0     & -5.7, +6.8  \\
\hline
Cross section           & 3.0   & 0.5     \\
\hline
Solar Model             & -16, +20  & -16, +20  \\
\hline
\end{tabular}
\end{center}
\end{table}

\begin{figure}[!p]
\begin{center}
\includegraphics[width=0.9\picwidth]{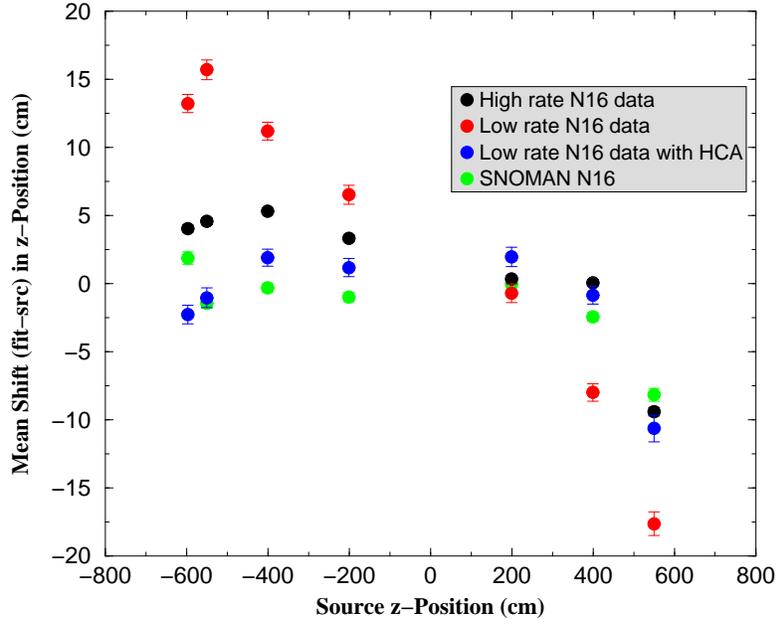}
\caption{The shift in the reconstructed position of the \N\ source as a function of
source position. The HCA calibration corrects the inward shift seen in the low rate
\N\ data. SNOMAN is the SNO Monte Carlo package.}
\label{recon}
\end{center}
\end{figure}

The systematic uncertainties in the SNO results are shown in table~\ref{sys_errs}. The
dominant uncertainties are the energy scale and the reconstruction accuracy. The
reconstruction accuracy was determined using a triggered \N\ 6.13\MeV\ $\gamma$-ray
source.~\cite{n16} Figure~\ref{recon} shows some of the results of such a study. When the
source was operated at low rate, the reconstruction accuracy was observed to become worse.
This was found to be because the PMT calibration characteristics were dependent on the readout
history of the PMT, compounded in non-central \N\ calibration runs by a readout rate gradient
across the detector. This was addressed by the HCA calibration~\cite{GMG} which allowed the
reconstruction accuracy of neutrino events to be correctly estimated at $\sim$3\% (rather than
$\sim$10\%).

\begin{figure}[!p]
\begin{center}
\includegraphics[width=0.9\picwidth]{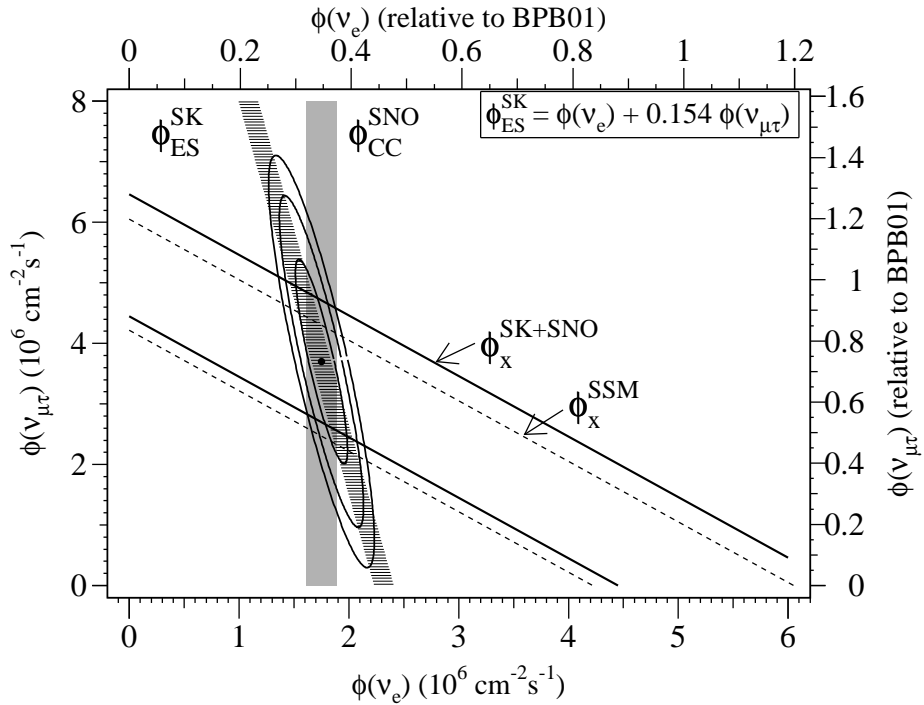}
\caption{The flux of \B\ solar neutrinos which are $\mu$ or $\tau$ flavour vs.\ the
flux of electron neutrinos as deduced from the SNO and Super-Kamiokande results. The
diagonal bands show the total \B\ flux as predicted by the BPB01 (dashed lines) and
that derived from the SNO and Super-Kamiokande results (solid lines). The intercepts
of these bands with the axes represent the $\pm$1$\sigma$ errors.}
\label{sno_res}
\end{center}
\end{figure}

\subsection{Total Active \iB\ Neutrino Flux}

Remembering that SNO's CC measurement is only sensitive to electron neutrinos,
whereas Super-Kamiokande's ES measurement has a weak sensitivity to all active
flavours, one can deduce the total \B\ solar neutrino flux. 
Stated explicitly, the experimental sensitivities to neutrino flavours are:
\[ \begin{array}{lclclcl}
\Phi^{\rm CC}_{\rm SNO} & = & \Phi_e \vspace{1ex} & ; &
\Phi^{\rm ES}_{\rm SK} & = & \epsilon\Phi_{\mu\tau} 
\end{array} \]
where $\Phi_{\mu\tau}$ is the combined $\nu_\mu$ and $\nu_\tau$ flux and
$\epsilon$=$1/6.481$. These equations can be solved for $\Phi_e$ and
$\Phi_{\mu\tau}$. This is shown graphically in figure~\ref{sno_res}.

The preferred value of the total active neutrino \B\ flux is:
\[ \Phi^{\rm TOT}_{\rm SNO+SK} = 5.44 \pm 0.99 \times 10^6 {\rm cm^{-2}\,s^{-1}} \]
which is in good agreement with the standard solar model prediction:
\[ \Phi^{\rm TOT}_{\rm BPB01} = 5.05^{+1.01}_{-0.81} \times 10^6 {\rm cm^{-2}\,s^{-1}} \]
This is the first determination of the total active flux of \B\ neutrinos generated
by the Sun.

\section{The NaCl Phase of the SNO Experiment}

The deployment of NaCl to enhance the NC capability of the SNO detector began
on May 28, 2001. The presence of NaCl in the \DO\ causes the free neutron,
produced by the NC interaction, to be captured by \Clf. This produces an
excited state of \Cls\ which decays to its ground state via a cascade of
$\gamma$-rays with a total energy of $\sim$8.6\MeV. The neutron detection
efficiency is significantly enhanced, and the high multiplicity of the
$\gamma$-ray cascade allows statistical separation from CC events based on the
PMT hit pattern.

\subsection{The \iNatf\ Calibration Source}

\begin{figure}[!t]
\begin{center}
\includegraphics[width=0.585\picwidth]{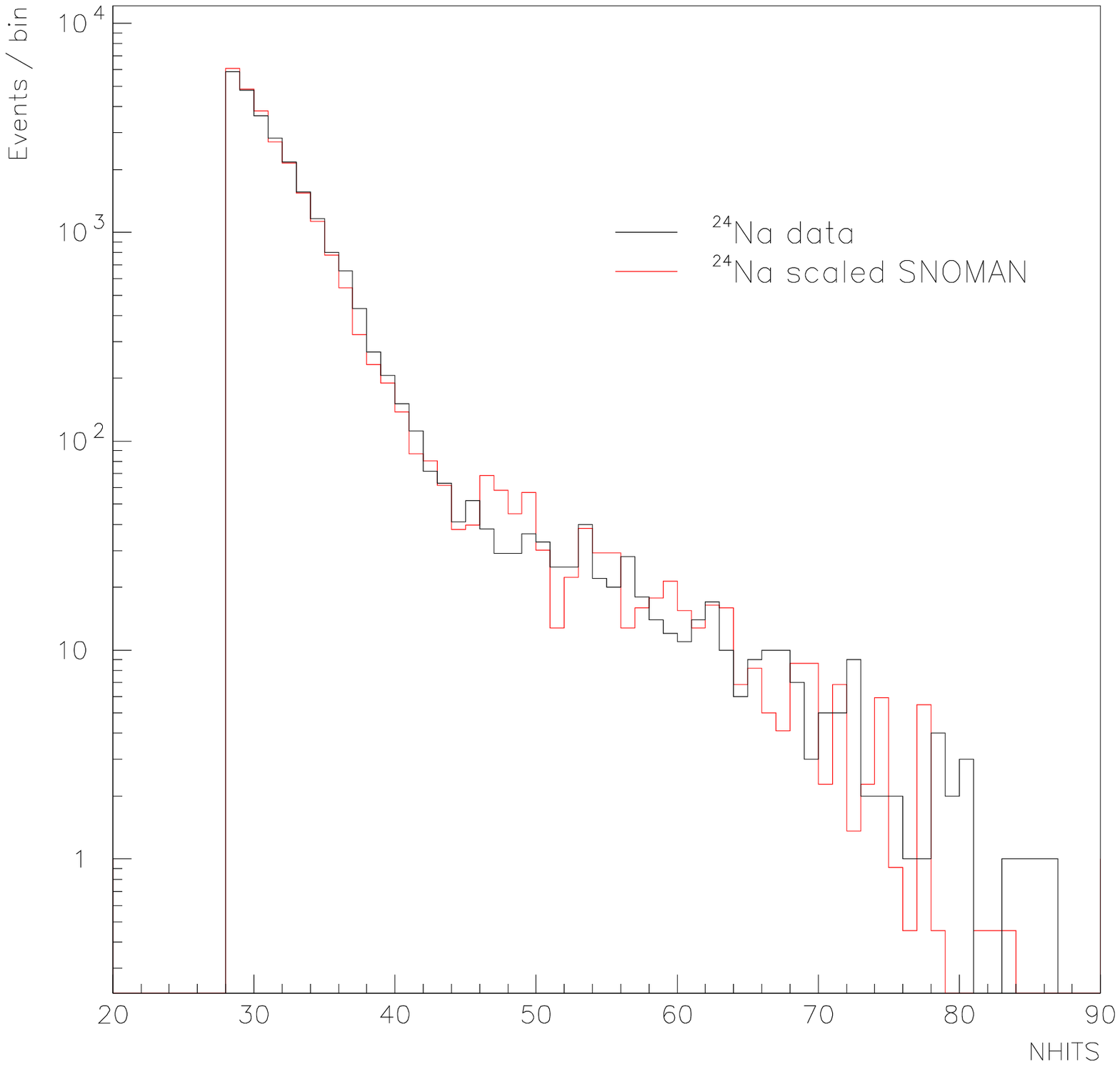}
\includegraphics[width=0.585\picwidth]{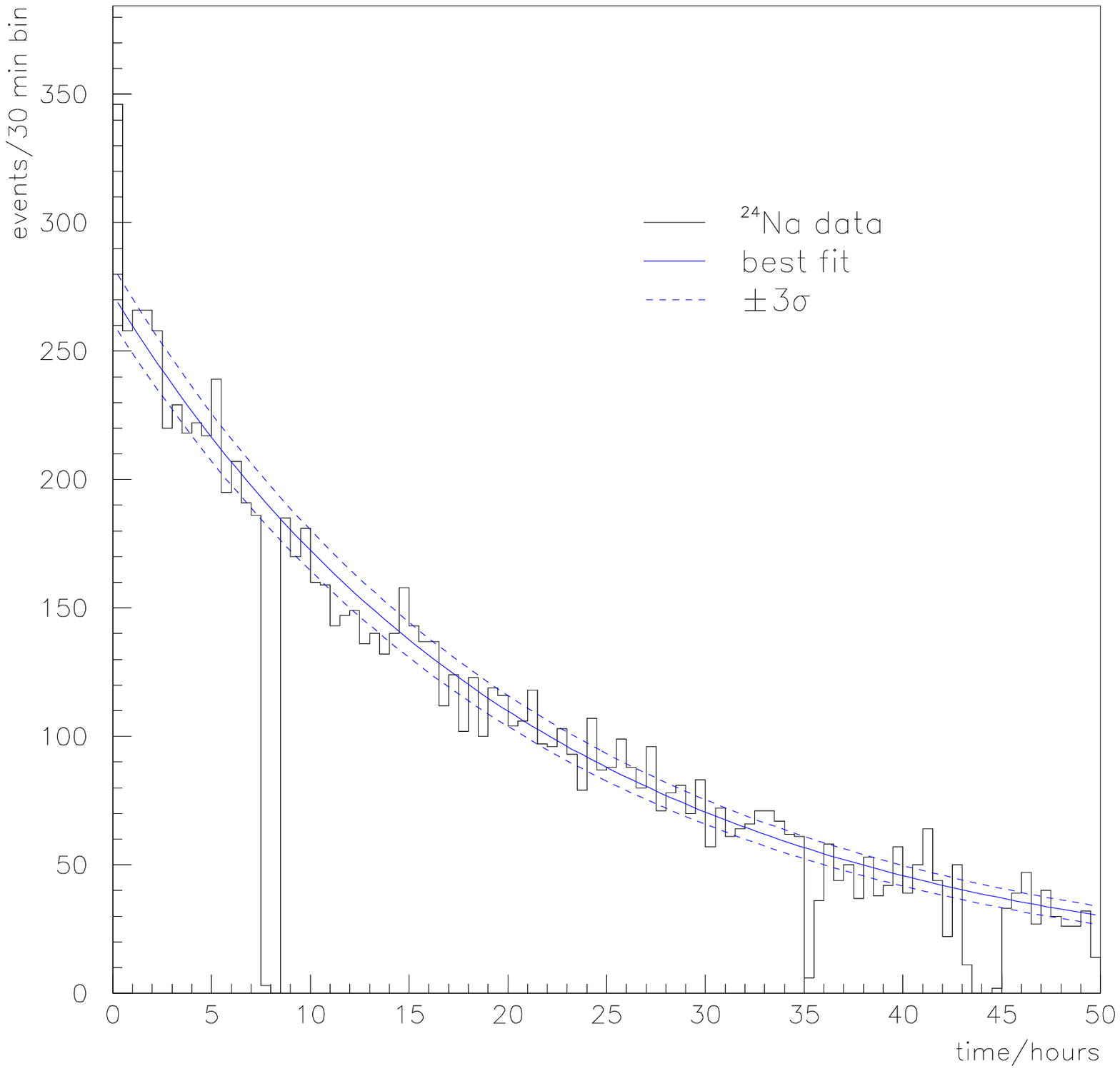}
\caption{Left: Comparison of the data and Monte Carlo NHITS distributions. The sharply falling
component at lower NHITS is from \Natf\ decays, and the higher NHIT bump is from neutron
capture on \Clf. Right: The decay of the activated \Natf\ in the \DO.}
\label{na24}
\end{center}
\end{figure}

The addition of NaCl to the \DO\ presented the opportunity to deploy \Natf\ as
a containerless source. This is desirable for two reasons: \Natf\ $\beta\gamma$
decays are similar to the $\beta\gamma$ decays of \Tl\ and \Bi; and a
containerless source avoids the difficulties in modeling complex sources.

Activating \Natt\ in the \DO\ was achieved by using the `super-hot' thorium
source, which produces 2.0$\times$10$^7$$\pm$5\% 2.614\MeV\ $\gamma$-rays per
minute (producing neutrons from deuteron photodisintegration). Figure~\ref{na24}
shows the results of such a deployment. Comparing the detector response from
the \Natf\ calibration source to Monte Carlo predictions gives confidence in,
and allows systematic uncertainties to be assigned to, techniques designed to
monitor the \Tl\ and \Bi\ levels within the \DO.

\section{Summary and Outlook}

Two significant results are reported in this paper. The data from SNO represents
the first direct evidence that there is an active non-electron flavour neutrino
component in the solar neutrino flux. This is also the first experimental
determination of the total flux of active \B\ neutrinos, which is in good
agreement with the solar model predictions.

The SNO Collaboration is now analysing the data from the pure \DO\ phase with a
lowered energy threshold. Efforts are devoted to understanding the low energy
$\beta\gamma$ decays of \Tl\ and \Bi\ and the photodisintegration contribution
they make to the NC measurement.

\section*{Acknowledgments}

This research was supported by the Natural Sciences and Engineering Research
Council of Canada, Industry Canada, National Research Council of Canada,
Northern Ontario Heritage Fund Corporation and the Province of Ontario,  the
United States Department of Energy, and in the United Kingdom by the Science
and Engineering Research Council and the Particle Physics and Astronomy
Research Council.  Further support was provided by INCO, Ltd., Atomic Energy of
Canada Limited (AECL), Agra-Monenco, Canatom, Canadian Microelectronics
Corporation, AT\&T Microelectronics, Northern Telecom and British Nuclear
Fuels, Ltd.   The heavy water was loaned by AECL with the cooperation of
Ontario Power Generation.

\end{document}